\documentclass{emulateapj}

\usepackage{amsmath}
\usepackage{natbib}
\usepackage[colorlinks=true,citecolor=blue,urlcolor=magenta,breaklinks]{hyperref}
\usepackage{cool}
\usepackage{eulervm}
\usepackage{times}

\begin{document}
\title[Neutron stars: A novel equation of state with induced surface tension]{Neutron stars: A novel equation of state with induced surface tension}
\author{Violetta V. Sagun~\altaffilmark{1,2},
	Il\'\i dio Lopes~\altaffilmark{1}}
\altaffiltext{1}{Centro Multidisciplinar de Astrof\'{\i}sica, Instituto Superior T\'ecnico, 
Universidade de Lisboa, Av. Rovisco Pais, 1049-001 Lisboa, Portugal; ilidio.lopes@tecnico.ulisboa.pt} 
\altaffiltext{2}{Bogolyubov Institute for Theoretical Physics, Metrologichna str. 14$^B$, Kiev 03680, Ukraine; violettasagun@tecnico.ulisboa.pt}

% % % % % % % % % % % % % % % % % % % % % % % % % % % % % % % % % % % % % % %
\date{\today}

\begin{abstract} 
 A novel equation of state with the surface tension induced by particles' interactions was generalized to describe the properties of the neutron stars (NSs). In this equation the interaction between particles occurs via the hard core repulsion by taking into account the proper volumes of particles.
 Recently, this model was successfully applied to the description of the properties of nuclear and hadron matter created in collisions of nucleons. The new approach is free of causality problems and is fully thermodynamically consistent, which enables us to use it for the investigation of the strongly interacting matter phase-diagram properties in a wide range of temperatures and baryon densities, including NSs. Here, we have calculated the mass-radius relations for a compact star using the Tolmann-Oppenheimer-Volkov (TOV) equation for two sets of parameters that satisfy the existing constraints. Accordingly, we found parameter values that are in good agreement with the same ones obtained from the nuclear-nuclear collision data analysis. The astrophysical constraints suggest that the hard core radius of baryons can vary between 0.425 fm and 0.476 fm. 
 
\end{abstract}

%\pacs{ln}

  
%%%%%%%%%%%%%%%%%%%%%%%%%%%%%%%%%%%%%%%%%%%%%%%%%%%%%%%%%%%%%%%%%%%%%%%%%%%%%
\section{Introduction}
%%%%%%%%%%%%%%%%%%%%%%%%%%%%%%%%%%%%%%%%%%%%%%%%%%%%%%%%%%%%%%%%%%%%%%%%%%%%% 

Neutron stars (NSs) are the most compact astrophysical objects in the universe that are accessible by direct observations ~\citep{2016ARA&A..54..401O}. Born during the gravitational collapse of luminous stars they provide a window into a world where the 
density of matter and strength of fields are much larger than those found in normal matter. Understanding their observed complex phenomena requires a wide range of scientific disciplines, including astrophysics, nuclear particle physics, and condensed matter physics. Knowledge of the interaction of particles in the superdense medium is particularly important for the formulation of a realistic equation of state (EoS) ~\citep{2015A&A...577A..40B, 2012ARNPS..62..485L}. The internal composition of the NS is poorly understood. It is thought that at the NS core the density can reach more than $ 10^{17} g~cm^{-3}$, while at the crust of the star it is even below the value of normal nuclear density $2.7 \cdot 10^{14} g~ cm^{-3}$ (which corresponds to the baryon density $n_{0}=0.16 fm^{-3}$).  At such densities the NS properties are determined predominantly by strong interaction between particles, mainly neutrons and protons. Therefore, modeling the NS interiors is important not only from the astrophysical but also from the particle physics point of view, since it opens up the possibility of studying the properties of the baryon matter under extreme conditions.

A reliable description of an NS is possible only with an account of the experimental data obtained within the recent nucleus-nucleus (A+A) collision programs. Despite the fact that in the terrestrial laboratories nuclear matter cannot be compressed to the densities achieved inside the NS, A+A collisions are an important source of information about the baryon matter EoS. Therefore, existing experimental data measured at AGS, SPS,  RHIC, and LHC facilities can be used to create a solid bridge between high-energy particle physics and astrophysics. For example, the study of transverse flows of particles through dense matter created in relativistic A+A collisions led to the formulation of constraints on the zero-temperature baryon matter EoS ~\citep{2002Sci...298.1592D}. Note that such an EoS is used as an input in EoS of NSs. By simultaneously taking into account the A+A collision experimental constraints and the astrophysical observational data from compact stars, such as neutron, quark, and hybrid stars, this gives us the unique opportunity to restrict the EoS inputs, consequently allowing us to perform high-quality modeling of stellar interiors in the high-density regimes of strongly interacting matter. Note that observational astrophysical data itself is able to discriminate between different nuclear matter EoSs. Thus, the highest-mass NS measurements~\citep[e.g.,][]{2010Natur.467.1081D,2013Sci...340..448A} provide the strongest observational constraints for the stiffness of the EoS at zero-temperature. Therefore, applying realistic EoSs that precariously reproduce the experimental data on A+A collisions to the description of NSs, guarantees a unified treatment of strongly interacting matter in a wide range of thermodynamic parameters of state.
 
A realistic EoS that is able to reproduce the properties of compact astrophysical objects has to fulfill several requirements. The possibility of including many particle species, which is known as multicomponent character, is of crucial importance for modeling the NS interiors, which in even the simplest treatment include neutrons, protons, and electrons, while more advanced descriptions have to account for the presence of hyperons \citep{2002PhRvL..89q1101S}. Therefore, the grand canonical ensemble is the natural choice for the formulation of such an EoS. Another element of the realistic phenomenological hadronic EoS corresponds to the short-range repulsive interaction of the hard core nature between particles ~\citep{2006NuPhA.772..167A, 2016arXiv161107349B}. Analysis of the particle yields produced in relativistic A+A collisions within statistical (thermal) models, i.e.,  the Hadron Resonance Gas (HRG) model \citep{2016arXiv161107349B}, shows the importance of the particle hard core repulsion. 
In this approach every particle species is defined as a rigid sphere with a fixed radius estimated from experimental data analysis. These radii do not exceed 0.5 fm ~\citep{2017JPhCS.779a2012A, 2017arXiv170300049S}. Note that the hard core of hadrons in phenomenological EoSs is important in order to suppress thermal excitations of the hadronic spectrum and provide deconfinement of the color degrees of freedom expected at high temperatures/densities ~\citep{2012LNP...841.....S}.
Another requirement to the phenomenological EoS is related to its causal behavior when the speed of sound cannot exceed the speed of light. At sufficiently high densities this condition is violated by the hard core repulsion. As it was shown by ~\citet{2017arXiv170300049S}, introducing the induced surface tension (IST) of particles to the model with the hard core repulsion between an arbitrary number of hadron species makes the EoS significantly softer and extends its causality range up to 7.5 normal nuclear densities, where formation of the quark-gluon plasma (QGP) is expected.
The IST is the key element of this approach \citep{2014NuPhA.924...24S}, as it allows us to account for the hard core repulsion between constituents in the most accurate way, and to properly reproduce the virial expansion of the multicomponent EoS. Recently, the IST EoS was used to describe the experimental data of hadron multiplicities measured at AGS, SPS, RHIC, and LHC energies of nuclear collisions~\citep{2017EPJWC.13709007S}, as well as the  nuclear matter properties~\citep{2014NuPhA.924...24S}. In this work the focus  is on the application of IST EoS to the study of NS properties.

One of the most important constraints on the realistic EoS of hadronic matter corresponds to the requirement that the properties of the nuclear matter ground state be reproduced ~\citep{1971ARNPS..21...93B}. Its stability requires a delicate balance between particle repulsion and attraction ~\citep{2012ARNPS..62..485L}. In this work, such an attraction is introduced within the mean field framework \citep{1988JPhG...14..191R} applied to IST EoS of quantum gases~\citep[see][for details]{2017arXiv170406846B}. This allows us not only to reproduce the properties of the normal nuclear matter but also to significantly extend the range of the IST EoS applicability. 

\smallskip

The goal of this work is to study the constraints that can possibly be imposed on the IST EoS from the astrophysical observational data of NSs. This is necessary to find which set of parameters of this EoS simultaneously fits the experimental data from the	A+A collisions and the astrophysical observations on NSs.
The paper is organized as follows. In Sec. \ref{sec-2} we briefly review the novel IST EoS and in Sect. \ref{sec-3} we present the mass-radius relation, including a discussion of constraints coming from nuclear  and high-energy nuclear physics experiments. The paper ends with a summary in Sect. \ref{sec-5}.~~

%%%%%%%%%%%%%%%%%%%%%%%%%%%%%%%%%%%%%%%%%%%%%%%%%%%%%%%%%%%%%%%%%%%%%%%%%%%%%
\section{Equation of state: model description}
\label{sec-2}
%%%%%%%%%%%%%%%%%%%%%%%%%%%%%%%%%%%%%%%%%%%%%%%%%%%%%%%%%%%%%%%%%%%%%%%%%%%%%

The novel thermodynamically consistent EoS is based on the virial expansion applied for the multicomponent mixtures with hard core repulsion ~\citep{2014NuPhA.924...24S}. In particular, proper consideration of the volumes of particles enables us to explicitly account for the surface tension that is induced by the interaction between constituents. In this formalism the IST EoS has the form of a system with two coupled equations, one for pressure $p$ and another for the IST coefficient $\sigma$ defined as follows:
\begin{eqnarray}
\label{Eq1}
p&=&\sum_ip_{i}^{id}(m_{i},~T, ~\mu_i-pV_i-\sigma S_i) \,
\end{eqnarray}
and
\begin{eqnarray}
\label{Eq2}
\sigma&=&\sum_iR_ip_{i}^{id}(m_{i},~T,~\mu_i-pV_i-\alpha\sigma S_i) \, ,
\end{eqnarray}
where $T$, $m_{i}$, $\mu_i$, and $R_i$  are temperature, mass, chemical potential, and the hard core radius of the $i$-sort of particles. The eigenvolumes and surfaces are equal to $V_i=\frac{4}{3}\pi R_i^3$ and 
$S_i=4\pi R_i^2$. The summations in Equations (\ref{Eq1}) and (\ref{Eq2}) are made over all types of particles and their antiparticles, with a corresponding partial pressure $p_i^{id}(m_{i},~T,~\nu_i)$ of point-like particles of type $i$. Here, $\nu_i$ denotes the shifted chemical potential due to the presence of the hard core repulsion. 

The dimensionless parameter $\alpha$ in Equation (\ref{Eq2}) was introduced due to the freedom of the Van der Waals extrapolation to high densities ~\citep{2014NuPhA.924...24S} and  accounts for the higher-order virial coefficients.  As was shown in ~\citet{2014NuPhA.924...24S} to reproduce the correct phase-diagram properties of nuclear matter, such a parameter should obey the inequality $\alpha>$ 1. According to the conditions of wide causality range and compressibility factor, the value of the parameter $\alpha$ in the Statistical Multifragmentation Model was found to be equal to 1.25 \citep{2014NuPhA.924...24S} . Further studies revealed the value $\alpha$=1.245 \citep{2017EPJWC.13709007S} is best for reproducing the fourth and third virial coefficients. 
Another physical meaning of $\alpha$ is as a switcher between the excluded volume and the proper volume regimes (for the details, see ~\citet{2017arXiv170300049S}). As the value $\alpha$=1.245 was obtained for the Boltzmann statistics ~\citep{2017arXiv170300049S} and could be different at zero-temperature, when quantum effects are significant,  the baryon radius and  $\alpha$ parameter are chosen to be free parameters in the model. On the other hand, when fitting the mass-radius curve to existing astrophysical constraints it is possible to extract the value of  $\alpha$ in a quantum case and give a prediction about how it depends on the thermodynamic parameters.

Originally, IST EoS (\ref{Eq1}) and (\ref{Eq2}) was written for classical  statistics to describe the nuclear multifragmentation ~\citep{2014NuPhA.924...24S} and hadron yields formed after the collision of heavy ions in the HRG model ~\citep{2017EPJWC.13709007S}. The fact that the model formulated in a grand canonical ensemble allows us to account for different particle species, i.e. hyperons and other baryons, which are required for a proper description of NS interiors with density $n > 2 n_{o}$ ($n_{0}=0.16 fm^{-3}$). Another  advantage of the IST EoS is related to its simple formulation in comparison to the standard formulation of the HRG model ~\citep{2012arXiv1204.0103O}. Thus, in the existing N-component statistical models it is required to solve N transcendental equations, which may include hundreds of corresponding hadronic species. In contrast to that, in the IST EoS the number of equations to be solved is 2 (Equations (\ref{Eq1}) and (\ref{Eq2})) and it does not depend on the number of hard core radii and particle species.

To apply an IST EoS to the description of the compact astrophysical objects at the vanishing temperature, the expressions for ideal pressure $p_{id}$ and density $n_{id}$ were rewritten for the Fermi gas particles with a quantum degeneracy factor $g=2$, due to two spin projections, as 
\begin{eqnarray}
\label{Eq3}
p_{id}(m,~\mu)&=&\frac{g \left(\mu k(2\mu^2-5m^2)+3m^4\ln\frac{\mu+k}{m}\right)}{48\pi^2} \,
\end{eqnarray}
and
\begin{eqnarray}
\label{Eq4}
 n _{id}(m,~\mu)&=&\frac{\partial p_{id}(m,\mu)}{\partial\mu}=\frac{g k^3}{6\pi^2}\,.
\end{eqnarray}
Here, $k=\sqrt{\mu^2-m^2}$ is the Fermi momentum, where  the chemical potential $\mu\ge m$.

To reproduce the NS properties of the IST EoS, the realistic interaction allowing one to go beyond the Van der Waals limit should be added. The phenomenological approach for how to add attraction between the constituents is related to the introduction of the additional density-dependent potential $U(n _{id})$ to the nucleon energy ~\citep{1988JPhG...14..191R} in the form of
\begin{equation}
\label{Eq5}
U( n _{id})=-C_d^2 n ^{\frac{1}{3}}_{id}\equiv-Ck\, ,
\end{equation}
where $C=C_d^2 \left(\frac{g}{6\pi^2} \right)^{\frac{1}{3}}$. 
Following ~\citet{1989ZPhyC..43..261B}, the power of density was chosen to be equal to $\frac{1}{3}$ to lead to the dimensionless $C_{d}^{2}$. Since the repulsion and attraction play the opposite roles, the $U(n _{id})$ term is introduced with a minus sign in the expression of the pressure of the system.

The condition of thermodynamic consistency of the model with the mean field interaction requires a special relation between the interaction pressure and potential ~\citep{1988JPhG...14..191R, 1989ZPhyC..43..261B, 2017arXiv170406846B}: 
\begin{eqnarray}
\frac{\partial p_{int}}{\partial n_{id}}&=&n_{id} \frac{\partial U(n_{id})}{\partial n_{id}}\,
\label{Eq6}
\end{eqnarray}
and
\begin{eqnarray}
p_{int}(n_{id}) = n _{id}U( n _{id})-\int_0^{ n _{id}}dn~U(n) = -\frac{gCk^4}{24\pi^2}\,. ~\
\label{Eq7}
\end{eqnarray}
Such a generalization of the EoS corresponds to the substitution of the pressure with $p(m,~\mu)\rightarrow p(m,~\mu-U(n _{id}))+p_{int}(n _{id})$. This parameterization provides the causal behavior of the model EoS at high densities. 

We considered a system that consists of neutrons, protons with the same value of hard core radius  $R_{n}=R_{p}=R_{n,p}$, and noninteracting point-like  electrons ($R_e$=0) with pressure $p_{id}(m_e,\mu_e)$, physical mass $m_e$, and chemical potential $\mu_e$.  Assumption about equal radii for neutrons and protons was justified in ~\citet{2016arXiv161107349B}. According to the previous findings, the best description of the experimental data corresponds to the baryon radius values lying in the range between $0.3$ fm and $0.5$ fm \citep{2016arXiv161107349B, 2017JPhCS.779a2012A, 2017arXiv170300049S}. The Coulomb interaction of electrically charged particles is neglected. Protons and neutrons, in addition to the interaction term (\ref{Eq7}), generate $p_{id}(m_p,\nu_p^1)$ and $p_{id}(m_n,\nu_n^1+\mu_e)$, where $m_p$ and $m_n$ are the proton and neutron masses and the shifted chemical potentials defined as
\begin{eqnarray}
\label{Eq8}
\nu^1_{A}&=&\mu_{A}-pV-\sigma S+U\Bigl(n^{id}(m_{A}, \nu^1_{A})\Bigl) \,
\end{eqnarray}
and
\begin{eqnarray}
\label{Eq9}
\nu^2_{A}&=&\mu_{A}-pV-\alpha\sigma S,
\end{eqnarray}
where $ A\in[p,n]$.	Here, $k_p^1=\sqrt{{\nu^1_p}^2-m_p^2}$ and $k_n^1=\sqrt{(\nu^1_n+\mu_e)^2-m_n^2}$ are the momenta of protons and neutrons,   correspondingly. Indexes $1$ and $2$ denote parameters from the Eq. (\ref{Eq1}) or (\ref{Eq2}). Thus, the system of equations for simultaneous determination of pressure $p$ and IST coefficient $\sigma$ in the present model can be written as
\begin{eqnarray}
\label{Eq10}
p&=&p_{id}(m_p,\nu_p^1)-\frac{gC{k_p^1}^4}{24\pi^2}+p_{id}(m_n,\nu_n^1+\mu_e)  \nonumber \\
&-&\frac{gC{k_n^1}^4}{24\pi^2}+p_{id}(m_e,\mu_e)\\
\label{Eq11}
\sigma&=&\left(p_{id}(m_p,\nu^2_p)+p_{id}(m_n,\nu^2_n+\mu_e)\right)R.
\end{eqnarray}
Note that electrons do not make any contribution to  Eq. (\ref{Eq11}) since they are treated in this 
work as point-like particles with $R_{e}=$0 fm. 

The EoS given by Equations (\ref{Eq10}) and (\ref{Eq11}) enables us to find particle number densities $n_i$ of neutrons, protons, and electrons as total derivatives of the pressure $p$ with respect to corresponding chemical potentials, i.e. $n_i=\frac{\partial p}{\partial \mu_i}$. Taking into account the condition of thermodynamic consistency given by Eq. (\ref{Eq6}), one gets
\begin{eqnarray}
\label{Eq12}
n_{A}&=&\frac{n^{id}(m_A,\nu^1_A)-n^{id}(m_A,\nu^2_A)\Phi}{1+\sum\limits_{A}n^{id}(m_A,\nu^1_A)V(1-\Phi)}\,
\end{eqnarray}
for $A\in [n,p]$ and
\begin{eqnarray}
\label{Eq13}
n_e&=&\frac{n^{id}(m_e,\mu_e)}{1+\sum\limits_{A}n^{id}(m_A,\nu^1_A)V(1-\Phi)}\,.
\end{eqnarray}
Here, the notation
\begin{eqnarray}
\label{Eq14}
\Phi=\frac{3\alpha\sum\limits_{A}n^{id}(m_A,\nu^2_A)V}{1+3\alpha\sum\limits_{A}n^{id}(m_A,\nu^2_A)V}.
\end{eqnarray}
is introduced. Baryon density $n_{B}$ can be found as the sum of neutron and proton particle densities, i.e.
\begin{equation}
\label{Eq15}
n_B=n_n+n_p
\end{equation}

To model the interior of an NS it is necessary to provide the electrical neutrality of the system, i.e. 
the zero value of the net density of the electric charge $n_Q$. The latter condition is equivalent to the
equality of particle densities of protons and electrons, which leads to
\begin{equation}
\label{Eq16}
n_Q=n_p-n_e=0\, .
\end{equation} 
This condition has to be accompanied by a balance of the particle chemical potentials due to the chemical equilibrium caused by equal rates of direct and inverse beta-decay processes $n\leftrightarrow p+e$. Note that neutrino contribution is neglected here. The neutron chemical potential is defined by the 
proton and electron chemical potentials as 
\begin{equation}
\label{Eq17}
\mu_n=\mu_p+\mu_e.
\end{equation}

The zero-temperature energy density of an electrically neutral equilibrated mixture of neutrons, protons, and electrons is defined by the thermodynamic identity
\begin{equation}
\label{Eq18}
\epsilon=\sum_{i=n,p,e}\mu_in_i-p=\mu_nn_B-p\, ,
\end{equation} 
where Equations (\ref{Eq16})-(\ref{Eq18}) are used on the second step. Expression (\ref{Eq18}) implicitly defines the dependence of the pressure on energy density, which is  needed in order to solve the TOV equation ~\citep{1934rtc..book.....T, 1939PhRv...55..364T, 1939PhRv...55..374O} in a closed form to model the internal structure of a compact star.

%%%%%%%%%%%%%%%%%%%%%%%%%%%%%%%%%%%%%%%%%%%%%%%%%%%%%%%%%%%%%%%%%%%%%%%%%%%%%
\section{Results and discussion}
\label{sec-3} 

After an EoS of NS matter is defined, the structure and properties of compact stars are obtained by solving the TOV equation for a spherically symmetric object of isotropic material that is in static gravitational equilibrium. Inserting an EoS into TOV equation, we create a connection between the internal properties of the star and its macroscopic properties. Integrating TOV equations from the center $r=0$ to the surface of the star $r=R$ where pressure $p=0$ and the total mass $m=M_{NS}$, we account for the boundary conditions. As NSs are electrically neutral, protons and electrons should be equal in number. 

As shown in Table \ref{tab1}, parameters that were taken from the description of the A+A collisions cannot reproduce the observed NS mass. A baryon radius lower than 0.4 fm makes the EoS very soft. This is due to the fact that adding the attraction term leads to a decrease of system pressure and the increase of the radii is necessary to reproduce the same data as that in the model without the attraction term. Therefore, the new sets of parameters were chosen (see Table \ref{tab1})  according to the criterion of reproduction of the astrophysical constraints, e.g. maximal observed NS mass  ~\citep{2013Sci...340..448A}, flow constraints taken from high-energy nuclear physics data ~\citep{2002Sci...298.1592D}, and a maximization of the causality range of the model. In addition, the value of the $C$ parameter was found according to the criterion of reproducing the properties of nuclear matter at the ground state.
The parameters $\alpha$ and $R_{n,p}$ of the IST EoS cause small 
effects on the nuclear matter ground state. The physical reason behind  this fact is related to the fact that at relatively low values of baryon density, the IST contribution is negligible. This specific point is discussed in  ~\citet{2014NuPhA.924...24S} and ~\citet{2017arXiv170406846B}, where these authors show that in this regime for $\alpha>1$ the present EoS turns to the Van der Waals one. Thus, the effective low-density approximation of the IST EoS does not include this parameter. As a result, all values of $\alpha$ that are used in the paper, do not affect the low-density limit. Furthermore, the contribution coming from the hard core radius of nucleons is also insignificant in this case, since in the grand canonical ensemble it enters the effective Van der Waals approximation of the IST EoS through the proper volume expression $V_{n,p}=\frac{4\pi}{3}R_{n,p}^3$, which comes multiplied 
by the pressure $p$ as $V_{n,p}p$ (see Equation (\ref{Eq1})). At the same time, the nuclear matter ground state pressure 
is zero, hence the contribution of the nucleon  proper volume $V_{n,p}$ is absent at this regime. This means
that for normal nuclear matter, the interparticle hard core repulsion is canceled by the mean filed attraction. Nevertheless, the effects of $\alpha$ and $R_{n,p}$ are principally at the high baryonic densities typical for NS interiors.

\begin{figure}[!]
\centering
\includegraphics[scale=0.4]{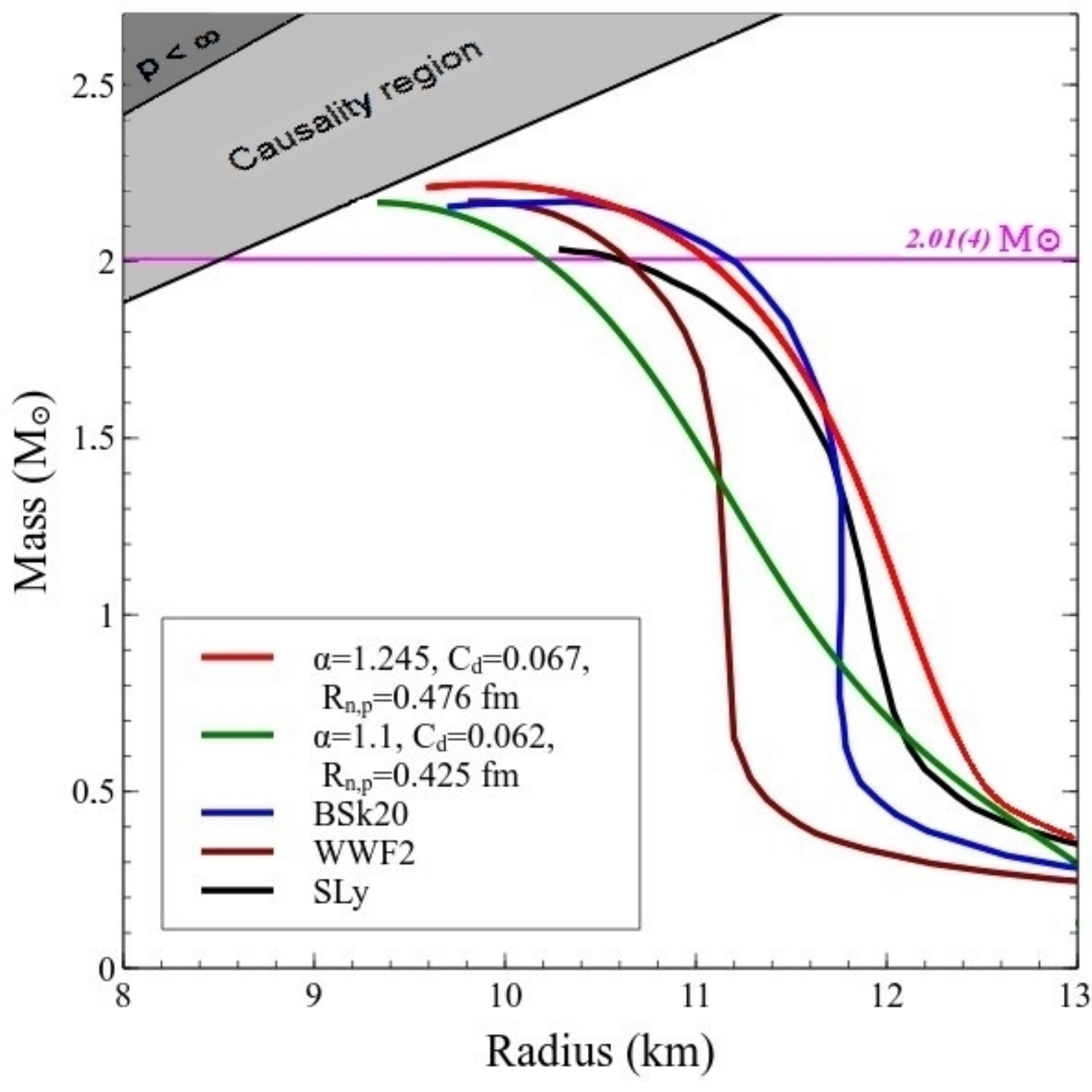}
\caption{
Gravitational mass-radius relation for IST EoS with two sets of model parameters:
the red curve represents the one that provides the best agreement with all astrophysical constraints, and the green curve corresponds to the limiting values of the model parameters that satisfy existing constraints. 
The additional curves correspond to the following EoS models: 
the blue curve represents BSk20~\citep{2013A&A...560A..48P};
the brown curve  corresponds to the WWF2 ~\citep{1988PhRvC..38.1010W};
and the black curve represents SLy ~\citep{2001A&A...380..151D}. 
The horizontal pink line defines the observational value of the most massive observed NS 2.01(4) $M_{\odot}$ ~\citep{2013Sci...340..448A}. The light gray region is excluded by causality constraint, $R >2.9GM/c^{2}$, and the dark gray region is excluded by the requirement of a finite pressure $R >2.25GM/c^{2}$  ~\citep{2007PhR...442..109L}.}
\label{fig:1}
\end{figure}

Figure~\ref{fig:1} presents a calculated mass-radius diagram for the IST model with $\alpha$=1.245, $R_{n,p}$=0.476 $fm$, and $C$=0.067 (red curve). These values for the parameters provide the best agreement with all astrophysical constraints. For this set of values, the maximum mass of static NS is 2.217 $M_{\odot}$ (see Table \ref{tab1}).

The values of $C$ for the green and red curves in Figure~\ref{fig:1} are almost the same, which is required by nuclear matter ground state properties. At the same time, even for such a small variation 
of $C$, the contribution is significant in the cases of large baryonic densities, which we found to be necessary to maintain in order for the IST EoS to accurately describe all the constraints coming from the astrophysical data. 
The above values were found by varying the model parameters -- $C$, $R_{n,p}$, and $\alpha$ -- in such way that their limiting values are able to describe the maximal observed NS mass, to provide a description of the nuclear matter properties at ground state, to define a maximal range of the model causality and to reproduce the flow constraints. The lower limit of the values of these parameters corresponds to the green curve in Figure~\ref{fig:1}, and the highest value allowable corresponds to the red curve in Figure~\ref{fig:1} (see also Table \ref{tab1}).
Thus, all the parameters that lead to models with values in the range between the green and red curves on Figure~\ref{fig:1} are in full agreement with all the existing constraints; as such, these values can be used to describe the properties of strongly interacting matter in NS and A+A collisions.

\begin{figure}[!]
\centering
\includegraphics[scale=0.4]{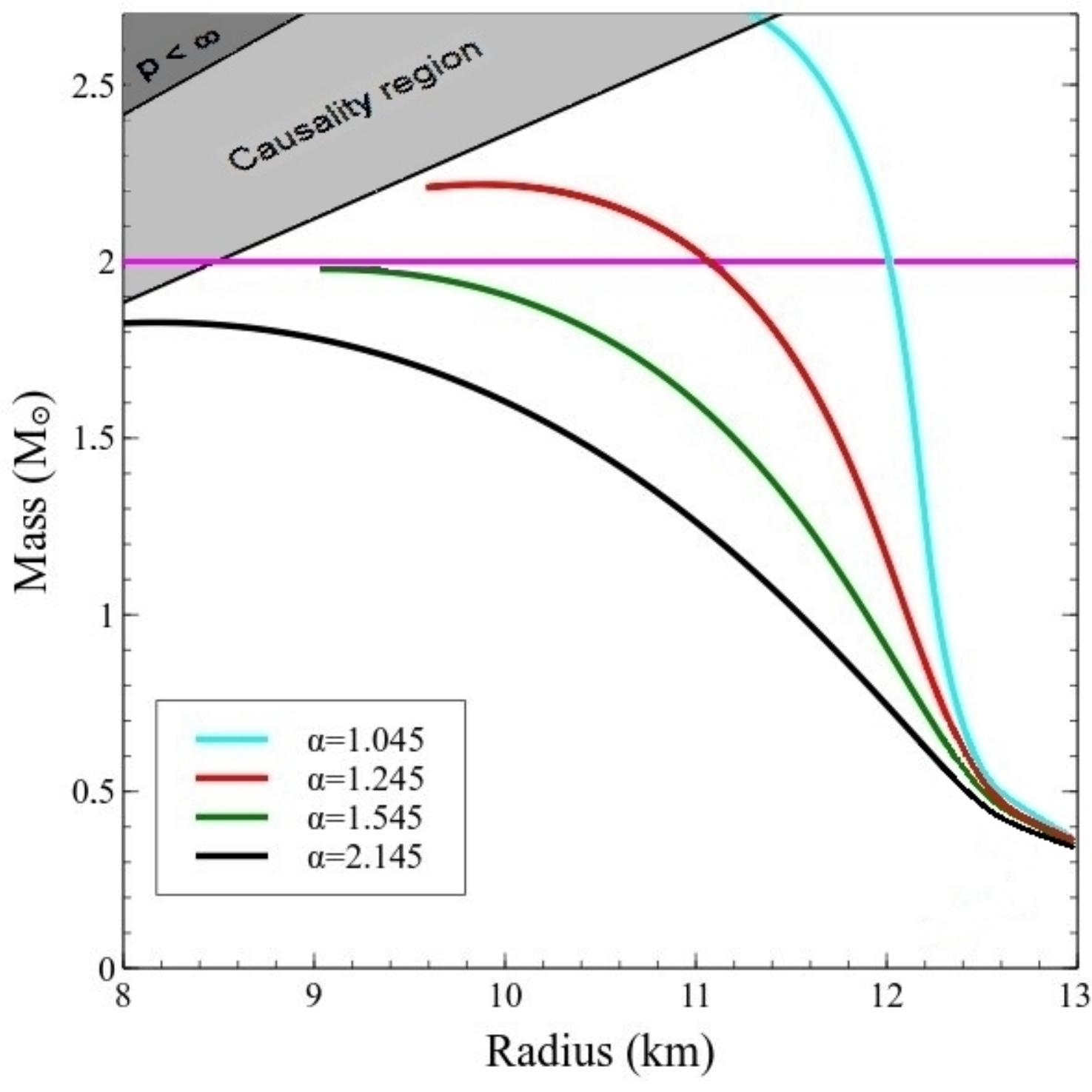} 
\includegraphics[scale=0.4]{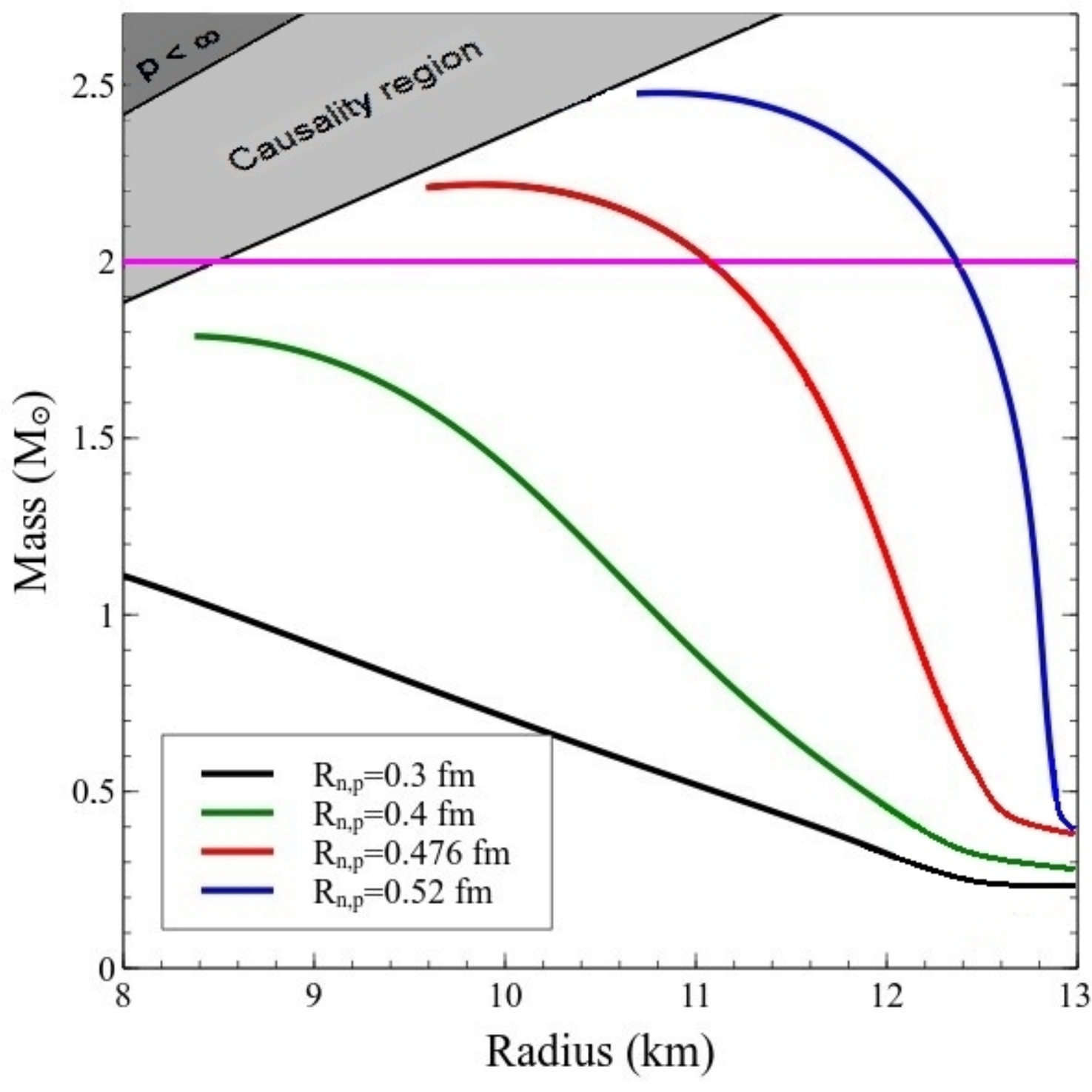} 
\caption{Effects of varying the $\alpha$ parameter (upper panel) and hard core radius of the baryons (lower panel). All other parameters are fixed to the values that correspond to the red curve in table \ref{tab1}.}
\label{fig:2}
\end{figure}

The present model is very sensitive to the hard core radius, which makes the EoS stiffer with its increase, and the $\alpha$ parameter contributes significantly at high densities, which affects the upper part of the mass-radius relation.
 As one can see from Figure~\ref{fig:2} (upper panel) the variation of $\alpha$ from 1.045 to 2.145 leads to a shift of the maximum of mass-radius curve. All other parameters were fixed to be equal to $R_{n,p}$=0.476 $fm$ and $C$=0.067. The value $\alpha$=1.045 does not satisfy the causality constraint depicted in light gray in Figure~\ref{fig:2}. The value of $\alpha$ that could reproduce the heaviest detected NS is  below 1.545.  The effect of varying the hard core radius of baryons is shown in Figure~\ref{fig:2} (lower panel). 
 To reproduce the observed compact stars the EoS should be quite stiff, but at the same time, a high $R_{n,p}$ value will lead to problems with causality, when the speed of sound could exceed the speed of light. The value $R_{n,p} \simeq$0.43 fm (all other parameters were fixed to the values from Table \ref{tab1} for curve A) is found to be the lowest one that could satisfy all astrophysical constraints.

\begin{table}[!]
\caption{\label{tab1} Values of the model parameters of the IST EoS found for the two limiting cases of neutron stars that satisfy all constraints and the parameters found from the description of A+A collisions  ~\citep{2017arXiv170300049S}. 
}
\hspace*{0.3cm}
\begin{center}
\begin{tabular}{|c|c|c|c|c|}
\hline
%model parameter &  \multicolumn{3}{c|}{value} \\
   & Baryon   & $\alpha$ & C  & Max\\
  &  Radius (fm)  &  &  & ${M_{NS}}/M_{\odot}$\\
\hline
Red curve &  0.476 & 1.245 & 0.067 & 2.217\\
in Figure ~(\ref{fig:1}) &  & &  & \\
\hline
Green curve &  0.425 & 1.06    & 0.062  & 2.166\\
in Figure ~(\ref{fig:1}) &  & &  & \\
\hline 
A+A collisions & 0.355    & 1.245 &  0.067 & 1.544\\
\hline   
\end{tabular}
\end{center}
\end{table}

\begin{figure}[!]
\centering
\includegraphics[scale=0.4]{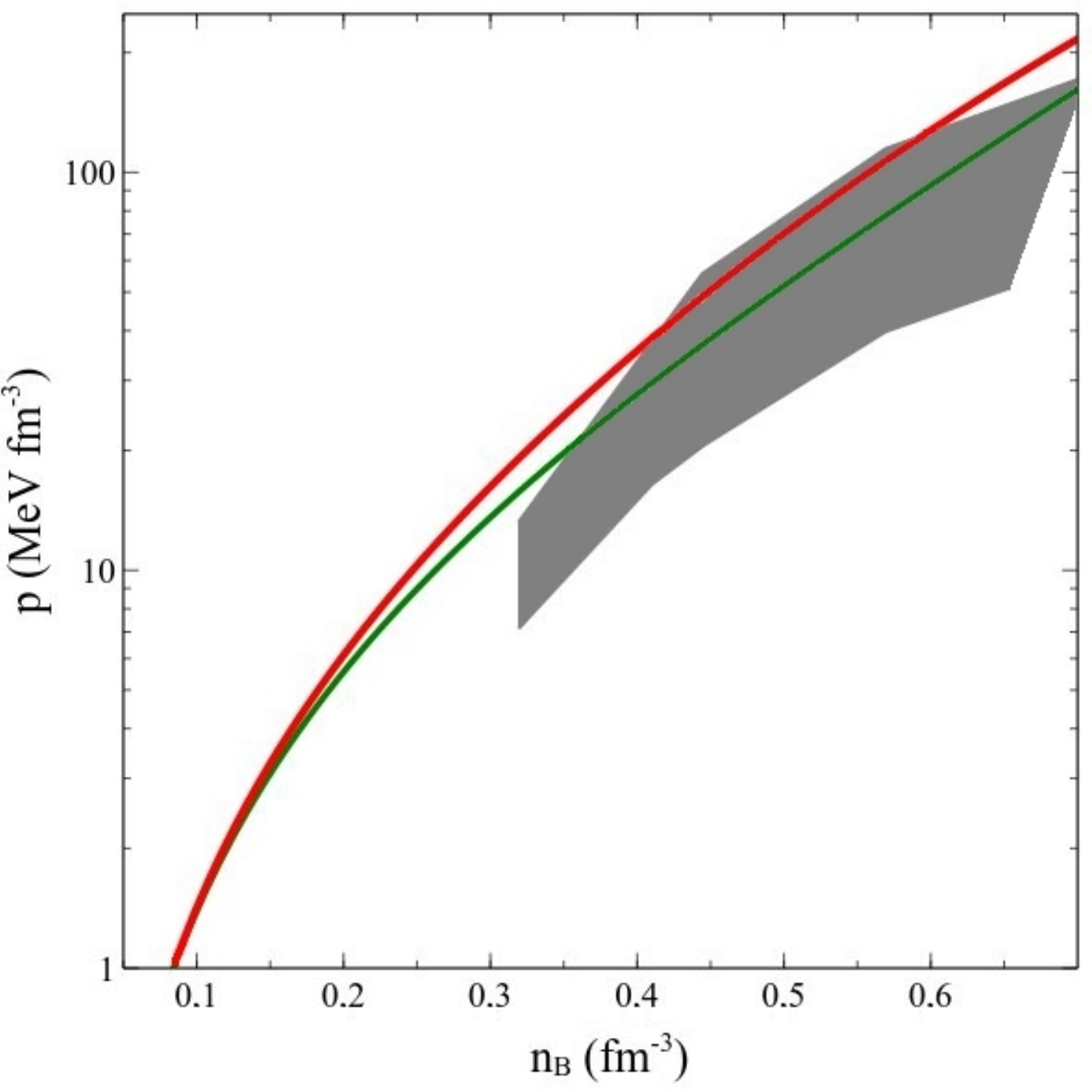} % Eigen functions
\caption{Dependence of the pressure $p$ on the baryon density $n_{B}$ for the defined set of parameters of the IST model. The red and green curves correspond to the curves of the same color in table \ref{tab1}.
}
\label{fig:3}
\end{figure}

The flow constraint taken from high-energy nuclear collisions for symmetric  baryon matter ~\citep{2002Sci...298.1592D} corresponds to the gray shaded area in Figure \ref{fig:3}. To satisfy it, the EoS is required to be rather soft at $n \sim 2n_{0}$ which puts very strong limitations on the existing model of strongly interacting matter. From Figure \ref{fig:3}, one can see that the area between the red and the green curves corresponds to the curves of the same color in Figure~\ref{fig:1}, which are in good agreement with the nuclear collisions.

\begin{figure}[!]
\centering
\vspace*{3mm}
\includegraphics[scale=0.4]{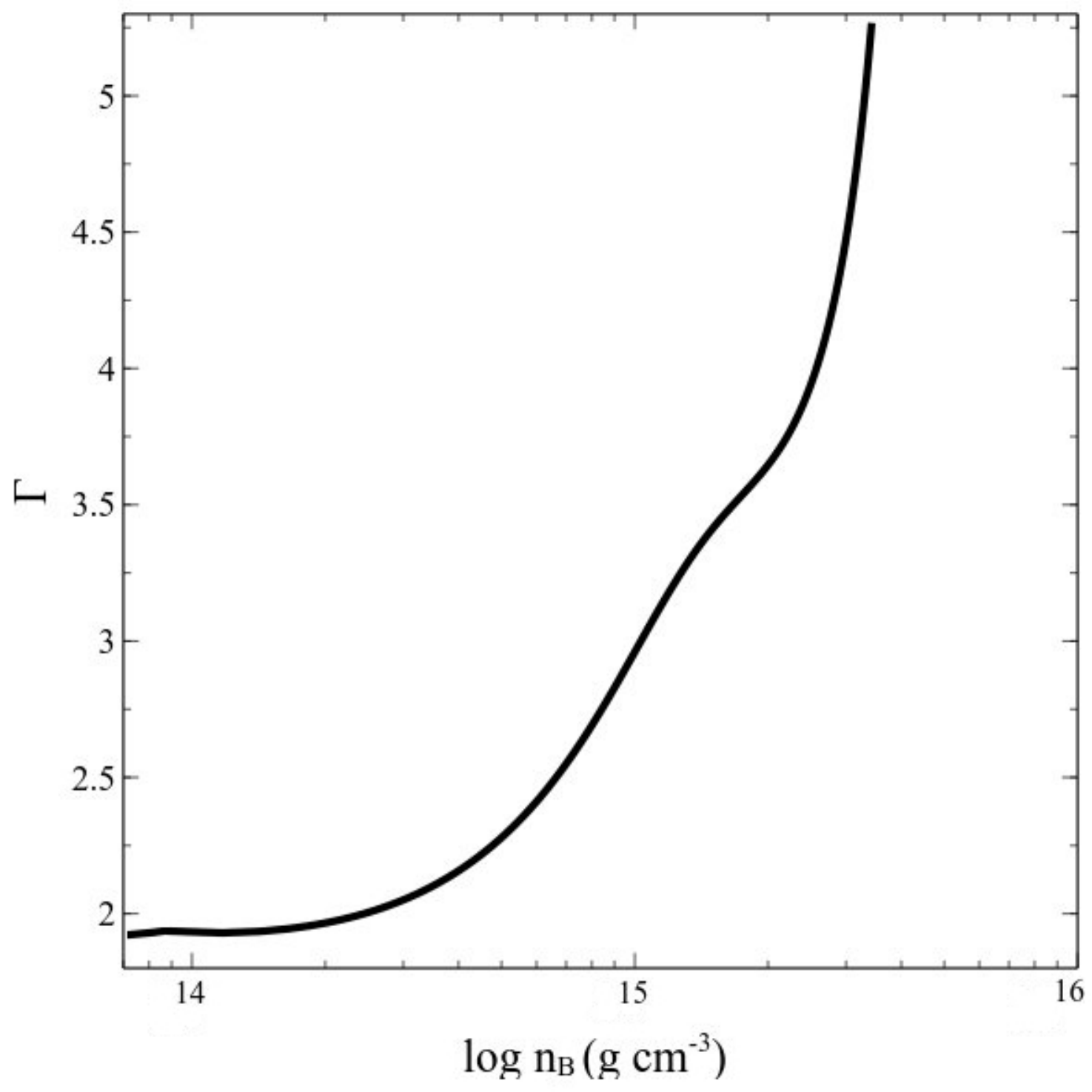} % Eigen functions
\caption{Adiabatic index of the IST EoS versus baryon density. 
}
\label{fig:4}
\end{figure}

One of the most prominent parameter that characterizes the physical properties of matter in different regions of the NS interior is the adiabatic index $\Gamma=\frac{n_{B}}{p}\frac{dp}{dn_{B}}$  ~\citep{2002A&A...394..213H, 2013A&A...560A..48P}. This dimensionless parameter determines the changes of pressure $p$ associated with the changes of the baryon density $n_{B}$ in the star. The value of adiabatic index characterizes the stiffness of the EoS at a given density and can shed light on the physical processes inside the NS. Analysis of $\Gamma$ allows us to define three layers inside the NS: outer zone, outer core, and inner core regions  ~\citep{2002A&A...394..213H}. In the outer zone region, which has the lowest density over  $n_{B}\sim 0.5 n_{0}$, the properties are almost independent of the density, while $\Gamma\simeq $ 1.7. The dependence of the adiabatic index on the baryon density is shown in
Figure ~\ref{fig:4}. It is seen that for densities higher than $7 \cdot 10^{13} ~g ~cm^{-3}$, where $\Gamma\simeq $ 1.75, the adiabatic index starts to monotonically increase. This behavior corresponds to compression of matter due to a transition to the outer core region of the NS. At $n_{B}\sim 2 n_{0}$ there is another transition to the inner core. In comparison to the RMF models, e.g. ~\citep{2013A&A...560A..48P}, which demonstrate a sudden jump of the adiabatic index $\Gamma$ at the transition region from the outer to inner core, the behavior of this parameter in the present EoS is more smoother.

%%%%%%%%%%%%%%%%%%%%%%%%%%%%%%%%%%%%%%%%%%%%%%%%%%%%%%%%%%%%%%%%%%%
\section*{Summary and Conclusions}
\label{sec-5}
%%%%%%%%%%%%%%%%%%%%%%%%%%%%%%%%%%%%%%%%%%%%%%%%%%%%%%%%%%%%%%%%%%%%%%%%%%%%%

Using a novel EoS within the IST between the constituents, we calculated the properties of an NS at the zero-temperature limit. The presented EoS is a thermodynamically self-consistent generalization of the IST model based on the virial expansion of the multicomponent mixture with hard core repulsion. Surface tension induced by particle interaction is a principally new element of that approach that enables us to go beyond the Van der Waals limit. It was shown that the present EoS can be successfully  applied to the description of the hadron multiplicities measured in A+A collisions, applied to studies of the nuclear matter phase-diagram, and applied to modeling of the NS interiors. Solving TOV equation, we obtained the mass-radius relations for the NSs. The values of the hard core radius found for baryons between 0.425 fm and 0.476 fm, $\alpha\in$ (1.06-1.245), and pressure-baryon density dependence  are in full agreement with A+A collisions. These results allow us to conclude that the description of the compact stars with the model, using  the description of the nuclear collision physics data,  provides a strong constraint on the attraction contribution in the EoS at zero-temperature. The IST EoS offers the possibility of describing the strongly interacting matter phase-diagram for a wide range of its thermodynamic parameters, which helps to create a solid bridge between the astrophysical and high-energy nuclear  physics data.

% % % % % % % % % % % % % % % % % % % % % % % % % % % % % % % % % % % % % % % % % %
\begin{acknowledgments}
V.S. is thankful for the fruitful discussions and suggestions from A.I. Ivanytskyi and K.A. Bugaev. The authors thank the Funda\c c\~ao para a Ci\^encia e Tecnologia (FCT), Portugal, for the financial support to the Multidisciplinary Center for Astrophysics (CENTRA),  Instituto Superior T\'ecnico,  Universidade de Lisboa,  through grant No. UID/FIS/00099/2013. The authors wish to thank the anonymous referee for the valuable comments and suggestions.
\end{acknowledgments}
  
\bibliographystyle{yahapj}

%------------------------------------------------------
%%%%%%%%%%%%%%%%%%%%%%%%%%%%%%%%%%%%%%%%%%%%%%%%%%%%%%%
%------------------------------------------------------
 
\end{document}